\pdfoutput=1

\documentclass[11pt]{article}

\usepackage[preprint]{acl}

\usepackage{times}
\usepackage{latexsym}

\usepackage{graphicx}
\usepackage{caption}
\usepackage{color}
\usepackage{amsmath}
\usepackage{booktabs}

\usepackage[T1]{fontenc}

\usepackage[utf8]{inputenc}

\usepackage{microtype}

\usepackage{inconsolata}

\usepackage{graphicx}

\title{A Nested Watermark for Large Language Models}

\author{
  Koichi Nagatsuka \\
  Hitachi, Ltd. \\
  \texttt{koichi.nagatsuka.mq@}\\\texttt{hitachi.com}
  \And
  Terufumi Morishita \\
  Hitachi, Ltd. \\
  \texttt{terufumi.morishita.wp@}\\\texttt{hitachi.com}\And
  Yasuhiro Sogawa \\
  Hitachi, Ltd. \\
  \texttt{yasuhiro.sogawa.tp@}\\\texttt{hitachi.com}
}

\begin{document}
\maketitle
\begin{abstract}
The rapid advancement of large language models (LLMs) has raised concerns regarding their potential misuse, particularly in generating fake news and misinformation. To address these risks, watermarking techniques for autoregressive language models have emerged as a promising means for detecting LLM-generated text. Existing methods typically embed a watermark by increasing the probabilities of tokens within a group selected according to a single secret key. However, this approach suffers from a critical limitation: if the key is leaked, it becomes impossible to trace the text’s provenance or attribute authorship. To overcome this vulnerability, we propose a novel nested watermarking scheme that embeds two distinct watermarks into the generated text using two independent keys. This design enables reliable authorship identification even in the event that one key is compromised. Experimental results demonstrate that our method achieves high detection accuracy for both watermarks while maintaining the fluency and overall quality of the generated text.

\end{abstract}

\section{Introduction}

Large language models (LLMs), such as GPT-4 \cite{achiam-2023-gpt4}, have achieved the ability to generate highly fluent text that is often indistinguishable from human-written content. As LLMs become increasingly widespread, they hold the potential to dramatically reduce the labor costs associated with tasks such as text composition, which have traditionally relied on human effort. However, concerns have also been raised regarding the misuse of LLMs for spreading misinformation and facilitating fraudulent activities \cite{crothers-2023-machine}. To address such issues, significant research has focused on developing techniques to detect text generated by LLMs.

Detection methods for LLM-generated text can be broadly categorized into (i) post-hoc detection and (ii) proactive (watermark-based) detection \cite{Kirchenbauer-2023-OnTR}. Post-hoc detection refers to methods applied after text generation, with typical approaches involving binary classification models that discriminate between human-written text and LLM-generated text \cite{jawahar-2020-automatic,mitchell-2023-detectgpt}. While these methods can be applied to outputs from arbitrary LLMs, their effectiveness fundamentally relies on a sufficient distributional gap between human- and machine-generated text. As LLMs continue to improve and this gap narrows, the performance of post-hoc detectors is expected to decline. In contrast, proactive detection methods embed detectable watermarks into the text during the generation process. Recent work has proposed watermarking techniques based on the autoregressive generation process of LLMs \cite{kirchenbauer-2023a-watermarkllm}, achieving substantially higher detection accuracy compared to post-hoc methods.

In watermarking for LLM outputs, a specific token pattern is embedded as a watermark by modifying the generation probabilities of a subset of tokens at each decoding step, according to a single secret key. For detection, the same secret key is used to identify the proportion of probability-adjusted tokens, and statistical hypothesis testing is performed to accurately determine the text's origin. However, a critical limitation of this approach is that, if the secret key used for watermarking is leaked, anyone with access to the key can reproduce the watermark, rendering reliable source attribution virtually impossible.

To mitigate this vulnerability to key leakage, this paper proposes a novel nested watermarking method. In the proposed approach, two distinct secret keys are used to embed two independent watermarks into the generated text. This enables robust source attribution even if the first key is compromised, as the second watermark remains secure and verifiable. Experiments conducted on instruction datasets constructed from prompt texts demonstrate that the proposed method achieves high detection accuracy for both watermarks. Furthermore, automatic evaluation experiments using GPT-4 confirm that our nested watermarking preserves the quality of generated text compared to single watermarking approaches.

\begin{figure*}[h]
    \centering
    \includegraphics[width=0.9\textwidth]{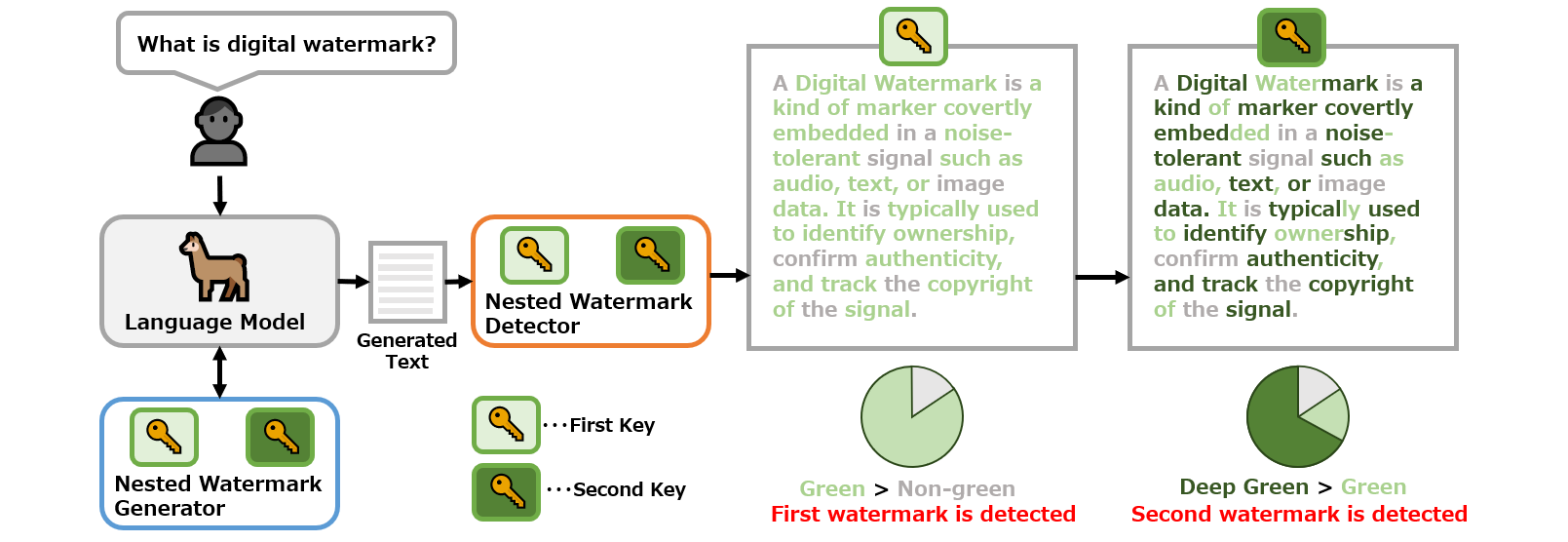}
    \caption{An overview of our nested watermark. The text on the right side of the figure demonstrates the detection of the first and second watermarks using the first and second keys, respectively. In the first text detected by the first key, the gray parts represent tokens classified as belonging to the token group without increased probabilities, while the light green parts indicate tokens classified as having increased probabilities. Furthermore, in the second text detected by the second key, the dark green parts signify tokens that belong to the group with increased probabilities during the embedding of the second watermark.}
    \label{fig:overview}
\end{figure*}

\section{Related Work}

Even before the advent of LLMs, research on embedding watermarks into text has been actively pursued for a considerable period \cite{kamaruddin-2018-review}. Early studies on text watermarking focused on exploiting the format and structure of textual data \cite{Brassil-1994-ElectronicMA}. Additionally, methods leveraging syntactic tree structures, which embed watermarks into text using keys, have also been proposed \cite{atallah-2001-natural}.

With the recent proliferation of LLMs, the need to distinguish between human-written and machine-generated text has grown rapidly, positioning watermarking as one of the most promising techniques for this purpose. A notable advantage of watermarking for LLM outputs is its empirically demonstrated robustness against text modifications \cite{Kirchenbauer-2023-OnTR}. On the other hand, vulnerabilities such as susceptibility to spoofing attacks have been identified, where adversarial edits are made to watermarked text without intentionally destroying the watermark itself \cite{sadasivan-2023-can}. Therefore, for watermarking and similar approaches to gain widespread adoption in real-world applications, it is essential to conduct ongoing research that considers a broader range of use cases.

\cite{zhu-2024-duwak} proposed Duwak, a dual watermarking scheme for large language models that embeds secret patterns in both the token probability distribution and sampling scheme using two keys, similar to our method; however, our approach is distinctive in that it does not require access to the model parameters in detection for the second watermark.

Beyond watermarking, another proactive detection approach is the use of retrieval-based systems \cite{krishna-2024-paraphrasing}. In these systems, outputs generated by LLMs are pre-registered in a database, allowing subsequent identification of the text's origin by reference to the stored entries. Although conceptually straightforward, this method is particularly effective as a countermeasure against the aforementioned spoofing attacks.



\section{Method}

Figure \ref{fig:overview} illustrates the overall architecture of the proposed method in the case of two nested watermarks. The framework comprises a nested watermark generator, a nested watermark detector, and multiple distinct secret keys. Within the nested watermark generator, watermarks are embedded at different layers using multiple keys as the language model generates text in response to a prompt. The nested watermark detector subsequently analyzes the output text and determines the presence or absence of each watermark by utilizing the corresponding keys. The following sections provide a detailed description of both the nested watermark generator and the nested watermark detector.

\subsection{Nested Watermark Generator}

Let $w_{t}$ be the $t$-th token in the text, and $p_{t}^{k}$ be the probability of the $k$-th token in the vocabulary $V$ at the $t$-th step. The probability $p_{t}^{k}$ is calculated using the softmax function:

\begin{equation}
p_{t}^{k} = \frac{\exp(l_{t}^{k})}{\sum_{i=1}^{|V|} \exp(l_{t}^{i})}
\end{equation}

where $l_{t}^{k}$ is the logit of the $k$-th token in the vocabulary $V$ at the $t$-th step.

We define a hash function, $H$, that maps the concatenation of the token $w_{t-n}$ at the $(t-n)$-th step and a secret key $s_{1}$ to a random number $r_{1}$, and the concatenation of a token $w_{t-m}$ at the $(t-m)$-th step and a secret key $s_{2}$ to a random number $r_{2}$, where $m \neq n$:

\begin{equation}
r_{1} = H(w_{t-n}, s_{1})
\end{equation}
\begin{equation}
r_{2} = H(w_{t-m}, s_{2})
\end{equation}

The random numbers $r_{1}$ and $r_{2}$ are used to determine the token groups $G_{1}$ and $G_{2}$, respectively. $G_{1}$ is a subset of the vocabulary $V$, and $G_{2}$ is a subset of $G_{1}$. The ratio of the size of $G_{1}$ to the size of $R_{1}$ (the set of remaining tokens in the vocabulary) is $\gamma : (1 - \gamma)$, where $\gamma$ is a hyperparameter.

To embed the watermarks, we add biases $\delta_{1}$ and $\delta_{2}$ to the logits of the tokens in $G_{1}$ and $G_{2}$, respectively. The total sum of the exponentiated logits, $D_{total}$, is calculated as follows:

\begin{equation}
\begin{split}
D_{total} = \sum_{i \in G_{1}, i \notin G_{2}} \exp(l_{t}^{i} + \delta_{1}) + \sum_{i \in R_{1}} \exp(l_{t}^{i}) \\ + \sum_{i \in G_{2}} \exp(l_{t}^{i}
+ \delta_{1} + \delta_{2})
\end{split}
\end{equation}

The adjusted probabilities for the tokens in $G_{1}$ and $G_{2}$ are then given by:

\begin{equation}
\hat{p}_{t}^{k} = \frac{\exp(l_{t}^{k} + \delta_{1})}{D_{total}}, \quad k \in G_{1}, k \notin G_{2}
\end{equation}
\begin{equation}
\hat{p}_{t}^{k} = \frac{\exp(l_{t}^{k} + \delta_{1} + \delta_{2})}{D_{total}}, \quad k \in G_{2}
\end{equation}

\subsection{Nested Watermark Detector}

To detect the presence of the watermarks ($G_{1}$ and $G_{2}$) in the text, we count the number of tokens belonging to $G_{1}$ and $G_{2}$, denoted $c_{1}$ and $c_{2}$, respectively. We then compute the z-scores $z_{1}$ and $z_{2}$ as follows:

For the first watermark:

\begin{equation}
z_{1} = \frac{c_{1} - \gamma T}{\sqrt{T \gamma (1 - \gamma)}}
\end{equation}

where $T$ is the total number of tokens in the text.

For the second watermark:

\begin{equation}
z_{2} = \frac{c_{2} - \gamma c_{1}}{\sqrt{c_{1} \gamma (1 - \gamma)}}
\end{equation}

If the z-scores $z_{1}$ and $z_{2}$ exceed a predetermined threshold $\theta$, we conclude that the corresponding watermarks are present in the text.






\begin{table*}
\begin{center}

\caption{Comparison of detection accuracy between nested watermark and single watermark}

\begin{tabular}{lcccc}

\toprule
{} & \multicolumn{2}{c}{\textbf{Nested watermark}} & \multicolumn{2}{c}{\textbf{Single watermark}} \\

\cmidrule(lr){2-3}
\cmidrule(lr){4-5}

& First watermark & Second watermark &  First watermark & First watermark \\

\toprule

bias ($\delta$) & 
1.5 & 
2.5 & 
1.5 & 
4.0 \\

\hline

z score (Avg.) & 
10.34 & 
7.94 & 
5.91 & 
12.52 \\

\hline

Type I error & 
0.000 & 
0.001 & 
0.000 & 
0.000 \\

Type II error & 
0.000 & 
0.012 & 
0.100 & 
0.000 \\

\hline

\label{tab:detection_performance}

\end{tabular}
\end{center}
\end{table*}

\section{Experiments}
\subsection{Experimental Setup}

\begin{figure}[t]
    \centering
    \includegraphics[width=0.40\textwidth]{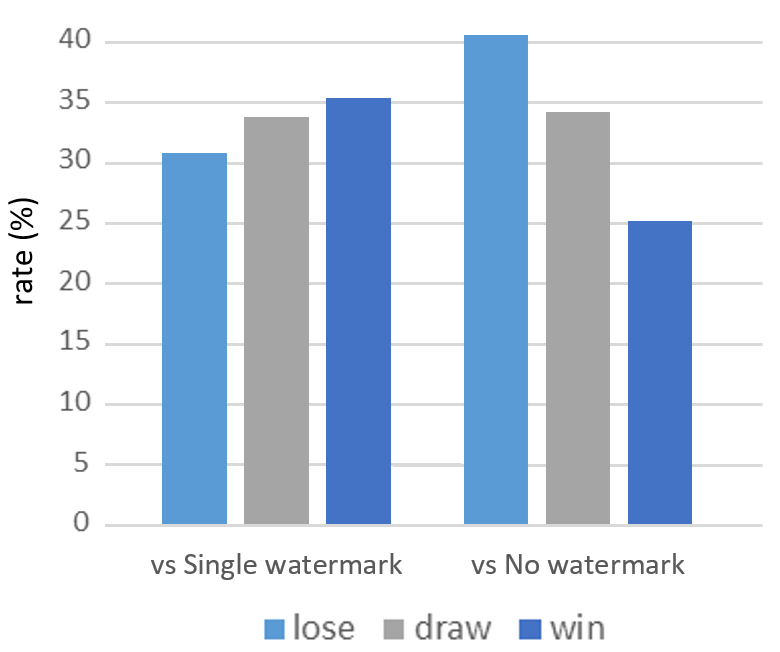}
    \caption{Comparison of text quality generated by each method. The vertical axis indicates the proportion of cases in which GPT-4 judged the text generated by the proposed method to be qualitatively superior (win), equivalent (draw), or inferior (lose) compared to the texts generated by each baseline.}
    \label{fig:text_quality_comparison}
\end{figure}

To evaluate the proposed method, we measure the detection accuracy of nested watermarking in terms of Type I error (false positives) and Type II error (false negatives). In addition, to assess the impact of nested watermark embedding on text quality, we conduct a quantitative evaluation using GPT-4 (gpt-4-1106-preview), following the LLM-as-a-judge framework \cite{zheng-2024-judging}, which is an automatic evaluation method. For both detection accuracy and text quality assessments, we compare the proposed method with a baseline of single-key watermarking (single watermarking) \cite{kirchenbauer-2023a-watermarkllm}. Notably, since LLM-as-a-judge is known to exhibit positional bias, where the text presented first tends to receive higher ratings, each comparison is conducted twice per example, swapping the order of the candidate texts, and the average score of both orders is reported.

For the evaluation dataset, we use 1,000 samples of English instruction data generated by gpt-4-1106-preview. This dataset consists of pseudo-prompts designed to reflect realistic use cases for LLMs (e.g., news articles and social media posts). In contrast, prior work \cite{kirchenbauer-2023a-watermarkllm} has focused on text completion tasks, where the prompts at inference time are constructed from fragmented texts sampled from C4 dataset. Therefore, utilizing our synthetic dataset allows us to evaluate the proposed method under a setting that is closer to real-world generation scenarios. 

We employ Llama-2-7b-chat as the text generation model. The maximum output token length is set to 300; if the generated text exceeds this limit, generation is stopped at 300 tokens. The ratio $\gamma$ of tokens with increased probabilities is set to 0.5 for embedding both the first and second watermarks. The increment applied to logits for nested watermarking is set to $\delta_1=1.5$ and $\delta_2=2.5$, respectively. The threshold for the z-score in statistical testing is set to 4.0. For decoding, beam search is used with a beam size of 8.

\begin{figure*}[t]
    \centering
    \includegraphics[width=\textwidth]{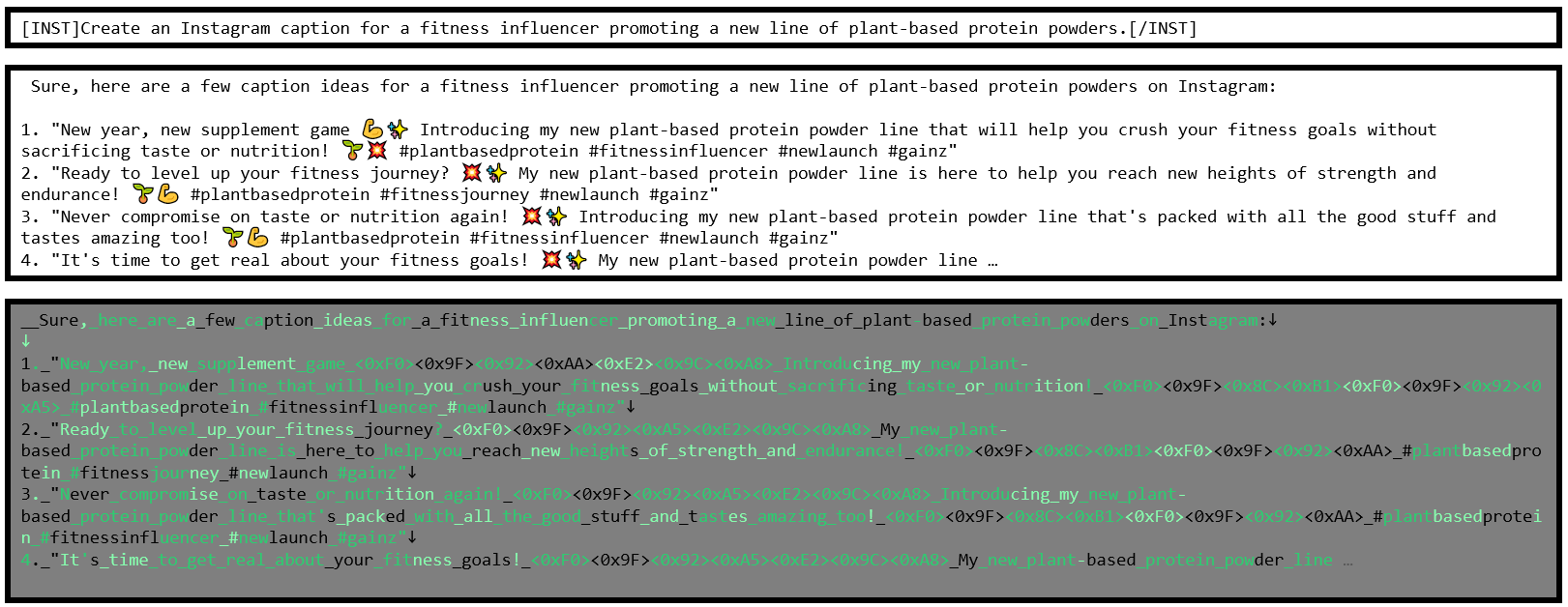}
    \caption{Example output of applying nested watermarking to Llama-2-7b-chat. The topmost box shows a prompt sampled from the instruction dataset. The second and third boxes display, respectively, the output text presented to the user and a visualization of the nested watermark in the text for detection purposes. In the third box, light green highlights indicate tokens marked by the first watermark, and dark green highlights indicate tokens marked by the second watermark. For readability, newline characters have been replaced with the symbol “↓” and line breaks have been processed accordingly.}
    \label{fig:output_example}
\end{figure*}

\subsection{Experimental Results}

\subsubsection{Detection Accuracy of Nested Watermarking}

Table \ref{tab:detection_performance} presents a comparison of detection accuracy between nested watermarking and single watermarking. For single watermarking, we evaluate two configurations: one where the bias is set equal to the first watermark in nested watermarking ($\delta_1=1.5$), and another where the bias equals the sum of $\delta_1=1.5$ and $\delta_2=2.5$, i.e., 4.0.

First, regarding the Type I Error (the proportion of cases where a watermark is incorrectly detected in non-watermarked text), all watermarking methods achieve an extremely low error rate of 0 or 0.1\%. In contrast, for Type II Error (the proportion of cases where a watermark is not detected in watermarked text), the error rate for single watermarking with a bias of 1.5 exceeds 10\%, indicating low detection performance. On the other hand, the Type II Error for the first watermark in the proposed method and for single watermarking with a bias of 4.0 is 0\%. As evidenced by the relatively high average z-scores for these methods, increasing the bias sufficiently raises the z-score in the statistical test, thereby reducing the Type II Error rate.

It is worth noting that in the proposed method, the token set for the first watermark includes the token set for the second watermark as a subset. Therefore, the bias for the intersection is the sum of $\delta_1=1.5$ and $\delta_2=2.5$, making it 4.0 in total, which leads to improved detection accuracy for the first watermark. Additionally, the Type II Error for the second watermark in the proposed method is 1.2\%. While there is room for improvement, this result confirms that the detection accuracy remains sufficiently high even in scenarios where the first key has been leaked.

\subsubsection{Text Quality of Watermarked Outputs }

Figure \ref{fig:text_quality_comparison} presents the results of the text quality comparison for each method. When comparing the proposed method to the no watermark baseline, the win rate is 25.2\%, while the lose rate is 40.6\%. This indicates that the introduction of nested watermarking tends to reduce the overall quality of text generated by the language model. However, the sum of the win rate and draw rate exceeds 60\%, suggesting that it remains reasonably feasible to generate fluent text even with nested watermarks applied.

Additionally, we conducted a comparison between the proposed method and the single watermark approach (single watermark) with a bias value of 4.0. Interestingly, relative to the single watermark, the proposed method achieved a higher win rate of 35.35\% and a lower lose rate of 30.85\%, indicating that our approach outperforms the single watermark method in terms of text quality. Intuitively, one might expect that a single watermark, which embeds a watermark only once, would have less impact on token generation probabilities compared to the nested watermarking approach, which embeds watermarks twice. However, considering the proportion of tokens affected by the bias, in the single watermark case, $\gamma \times |V|$ tokens are uniformly affected by a bias of 4.0. In contrast, for nested watermarking, $\gamma \times |V|$ tokens are first influenced by the initial bias (1.5), but only $\gamma^2 \times |V|$ tokens are subsequently affected by the second bias (2.5). Therefore, the number of tokens influenced by the maximum bias (4.0) is reduced by a factor of $\gamma$ compared to the single watermark case. This difference likely contributes to the improved preservation of text quality observed with the proposed method.

\section{Conclusion}

In this paper, we proposed a nested watermarking scheme that enables source attribution even in cases where the first key is leaked. Experimental results demonstrated that the proposed method achieves high detection accuracy for both watermarks from two keys, as measured by Type I and Type II Errors. Furthermore, compared to conventional methods, our approach maintains text quality at an equal or higher level. Future work includes investigating the robustness of nested watermarking against text tampering and evaluating detection accuracy on shorter texts.

\bibliographystyle{acl_natbib}

\bibliography{custom.bib}

\end{document}